# Effect of Sintering Temperature on Structural and Magnetic Properties of $Ni_{0.6}Zn_{0.4}Fe_2O_4$ Ferrite: Synthesized from Nanocrystalline Powders

M. A. Ali[1*], M. N. I. Khan[2], F. -U. -Z. Chowdhury[1], D. K. Saha[2], S. M. Hoque[2], S. I. Liba[2], S. Akhter[2], and M. M. Uddin[1]

[1]*Department of Physics, Chittagong University of Engineering and Technology (CUET), Chittagong-4349, Bangladesh.*
[2]*Materials Science Division, Atomic Energy Center, Dhaka-1000, Bangladesh.*

*Email: ashrafphy31@gmail.com

**Abstract:**

The effect of sintering temperatures ($T_s$) on the structural and magnetic properties of $Ni_{0.6}Zn_{0.4}Fe_2O_4$ (NZFO) ferrites synthesized by conventional double sintering method has been reported. The samples are sintered at 1200, 1250 and 1300 °C. The X-ray diffraction (XRD) analysis reveals the formation of a single phase cubic spinel structure of the sample. The magnetic parameters such as saturation magnetization, $M_s$; coercive field, $H_c$; remanent magnetization, $M_r$ and Bohr magneton, $\mu_B$ are determined and well compared with reported values. The obtained values are found to be 71.94 emu/gm and 1.2 Oe for $M_s$ and $H_c$, respectively at $T_s$=1300 °C. Curie temperature ($T_c$) at various $T_s$ has also been calculated. It is noteworthy to note that the sample with a very low $H_c$ could be used in transformer core and inductor applications.

## 1. Introduction

Nickel–Zinc ferrites have attracted huge attention in recent years owing to their potential applications in power transformers in electronics, antenna rods, loading coils, microwave devices and telecommunication, and are promising candidates for microwave device applications due to their remarkable magnetic properties, high resistivity and low eddy current loss [1-5]. The properties of the ferrites sensitively depend on the method of preparation, dimension of raw materials (nano or bulk), composition, sintering conditions [6, 7]. Several authors have reported the structural, electrical and magnetic properties of the Ni-Zn ferrites prepared from bulk or nano powder [8-18]. It is expected that dimension of the raw materials (nano powder) could be changed the magnetic as well as electrical properties of the Ni-Zn ferrites. However, most cases, the Ni-Zn ferrites have been prepared from bulk powders or claimed synthesized nanocrystalline powders where the crystallite size are calculated using Scherrer equation by analyzing X-ray diffraction (XRD) peaks. It is noted that the determination of the crystallite size by the Scherrer equation has some limitation for solid state synthesis. Therefore, it is great importance to study the effect of sintering temperature ($T_s$) on the properties of Ni-Zn ferrite synthesized by nano powders as raw materials.

In this work, we have studied the effect of $T_s$ on the structural and magnetic properties of NZFO prepared by conventional double sintering route.

## 2. Material and Methods

The polycrystalline samples of NZFO were synthesized using conventional double sintering techniques route by taking reagent grade oxides of nickel, zinc and iron nano powders of purity greater than 99.5% (US Research Nanomaterials, Inc.). The particle sizes of $Fe_2O_3$, NiO, and ZnO nano powders are ~ 20-40, 15-35 and 35-45 nm, respectively were weighed according to the corresponding composition. The weighed powder was mixed and ground for 3 hrs using an agate mortar and pestle. The slurry was dried and loosely pressed into cake using a hydraulic press. The cake is pre-sintered in air for 3 hrs at 900°C. The pre-sintered cake removed from the furnace was crushed and again ground for 1 hr. The obtained powders was then pressed using a suitable die in the form of ring (11 mm diameter; 3.4 mm thickness) with a hydraulic press at a pressure of 15 kN using 5% PVA solution as a binder and the samples were finally sintered at 1200, 1250, and 1300°C for 4 hrs in air. The crystalline phases of the prepared samples were studied by XRD [Philips X'pert Pro X-ray diffractometer (PW3040)] with Cu-$K_α$ radiation ($λ$ =1.5405 Å). The magnetic properties (*M-H* curve, saturation magnetization, $M_s$; coercive field, $H_c$; remanent magnetization, $M_r$; and Bohr magneton, $μ_B$) were determined by the vibrating sample magnetometer (VSM) (Micro Sence EV9) with a maximum applied field of 15 kOe. Frequency and temperature dependent permeability were investigated by using Wayne Kerr precision impedance analyzer (Model 65120B).

## 3. Results and Discussion

### 3.1. Structural properties

Fig. 1 shows the powder XRD pattern of the calcined NZFO at 900°C. Very sharp, broad and well-defined peaks have been observed from the XRD pattern indicating the cubic spinel phase of NZFO. The average crystallite size has been calculated which is found to be 51 nm [19].

Single phase cubic spinel structure with $Fd\bar{3}m$ space group symmetry of NZFO has also been confirmed at different $T_s$ by the XRD study (data not shown here) [20].The average grain sizes are determined from the SEM micrograph by linear intercept technique and ranging between 1.4 - 7.8 µm, which are almost homogeneously distributed throughout the sample surface [20]. The obtained average grain sizeare significantly smaller than that of ref. [18] (size ranging between 8-17 µm) for NZFO ferrites synthesized from bulk powders sintered at 1200, 1250, and 1300°C. Smaller grains have a meaningful influence on the magnetic as well as electrical properties of NZFO ceramics.

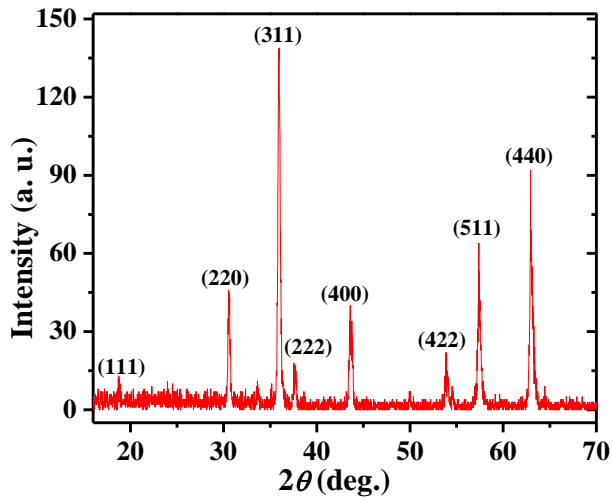

**Fig. 1.** The XRD pattern of $Ni_{0.6}Zn_{0.4}Fe_2O_4$ ferrites calcined at 900°C for 3 hrs in air.

*3.2. Magnetic properties*

The room temperature static applied magnetic field, H (up to 15 kOe) magnetization hysteresis loops of the NZFO ceramics sintered at various temperature 1200, 1250, and 1300 °C is shown in Fig. 2. The value of magnetization increases with increasing applied magnetic field up to a certain field above which the sample becomes saturated. It is seen that the values of coercive field ($H_c$) of the samples are found to be quite low < 3 Oe indicating the studied samples are soft ferrites. Moreover, our obtained values are also smaller than that obtained by Hossain et al. [18]. The saturation magnetization ($M_s$), coercive field ($H_c$), remanent magnetization, $M_r$, and Bohr Magneton, $\mu_B$, are calculated from the measured magnetic hysteresis loop and are presented in Table 1. It is depicted that the value of $M_s$ increases with increasing $T_s$ (Fig. 2 (b)) while the value of $H_c$ decreases. This typical behavior can be attributed taking into account variation of grain size that is strongly dependent on the $T_s$. It is observed at low $T_s$, the grain size is smaller than that of higher $T_s$. It is obvious that a large grain contains numerous domain walls. In addition, the number of domain walls also increases as the crystallite sizes increases with the increasing $T_s$. Thus, the domain wall motion is affected by the grain size and enhanced with the increase of grain size [21].

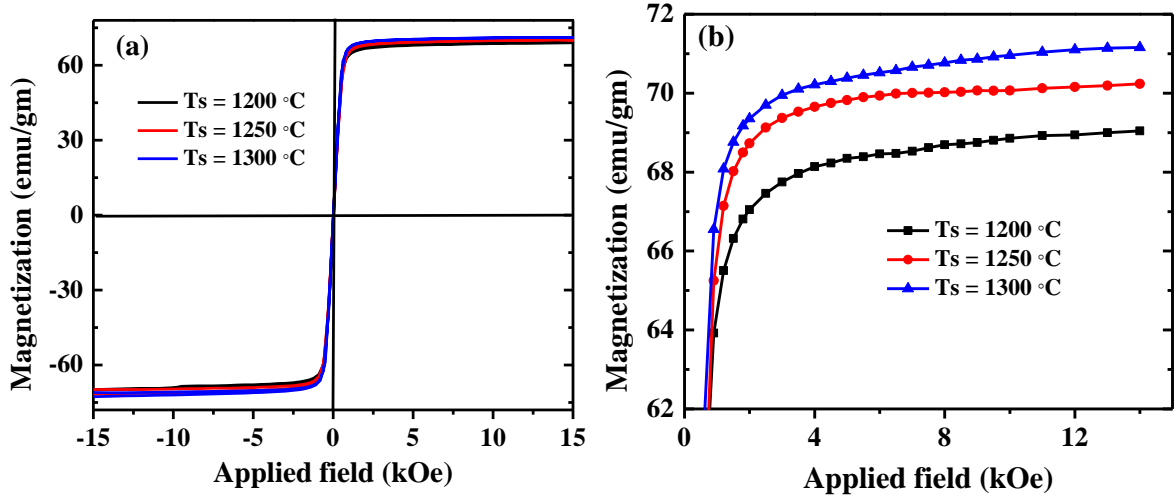

**Fig. 2.** The $M$–$H$ loops of $Ni_{0.6}Zn_{0.4}Fe_2O_4$ ferrite samples sintered at 1200, 1250 and 1300°C.

**Table 1.** The parameters of Curie temperature, $T_c$, saturation magnetization, $M_s$, coercive field, $H_c$, remanent magnetization, $M_r$ and Bohr magneton, $n_B$ in comparison with reported value for different $T_s$.

| Sintering Temperature, $T_s$ (°C) | Curie Temperature, $T_c$ (°C) | $M_s$ (emu/gm) | $H_c$ (Oe) | $M_r$ (emu/gm) | $\mu_B$ (Bohr magnetron) |
|---|---|---|---|---|---|
| 1200 | 190 | 69.4272 | 2.6720 | 0.33 | 2.94 |
|  |  | 72.9 [16] |  |  |  |
| 1250 | 215 | 70.9906 | 2.3460 | 0.28 | 3.01 |
|  | 350 [18] |  |  |  |  |
| 1300 | 240 | 71.9468 | 1.2590 | 0.16 | 3.05 |
|  |  | 79 [18] |  |  |  |
|  |  | 72.2 [16] |  |  |  |

Domain wall needs low energy than that of domain rotation to produce magnetization. Therefore, the contribution of domain wall movement in the magnetization is greater than that of domain rotation. It is also seen that the $M_s$ increases while the $H_c$ decreases with increasing $T_s$ which can be elucidated by the Brown's relation:, $H_c = 2K_1/\mu_0 M_s$, where $K_1$ is the anisotropy constant and $\mu_0$ is the permeability of free space. Furthermore, the $M_r$ decreases while the $\mu_B$ increases with increasing $T_s$, as expected for ferromagnetic materials.

Frequency dependent real part of the initial permeability ($\mu'$) at various $T_s$ is shown in Fig. 3 (a). It is noticed that the $\mu'$ is fairly constant up to certain low frequencies with maximum (214 at 4 MHz for $T_s$=1300°C and then falls rather rapidly to a very low value at high frequency (100 MHz). The fairly constant $\mu'$ values with a wide range frequency region is known as the zone of utility of the ferrite that demonstrate the compositional stability and quality of ferrites prepared by conventional double sintering route. This characteristic is anticipated for various applications such as broadband pulse transformer and wide band read-write heads for video recording [22]. At higher $T_s$, the permeability value is higher and the frequency of the onset of ferrimagnetic resonance is lower that is good agreement with Snoek's limit $f_r \mu'_i$ = constant [23], where $f_r$ is the resonance frequency of domain wall motion, above which $\mu'_i$ decreases. The ferrite materials exhibit higher permeability but due to Snoek's limit, have to be used according to the frequency range in use. It is also seen that the value of $\mu'$ increases with increasing $T_s$ (grain size and $M_s$ increases with increasing $T_s$ as discussed in previous section and table 1), and follows the Globus relation $\mu_i \propto \frac{M_s^2 D}{\sqrt{K_1}}$ where $M_s$ is the saturation magnetization, $D$ is the average grain size and $K_1$ is the magneto-crystalline anisotropy constant, as we expect for ferrite materials. The stability region of NZFO is found to be ~ 1kHz to 1.5 MHz, which is greater than that of NZFO synthesized from bulk raw materials [18]. It is also observed that the permeability increases with increasing $T_s$ while the utility zone decreases, larger grain growth at higher $T_s$ could be attributed this trend. Moreover, at higher $T_s$, the density is higher and larger grain size, near and accelerates grain to grain continuity in magnetic flux leading to higher permeability [22].

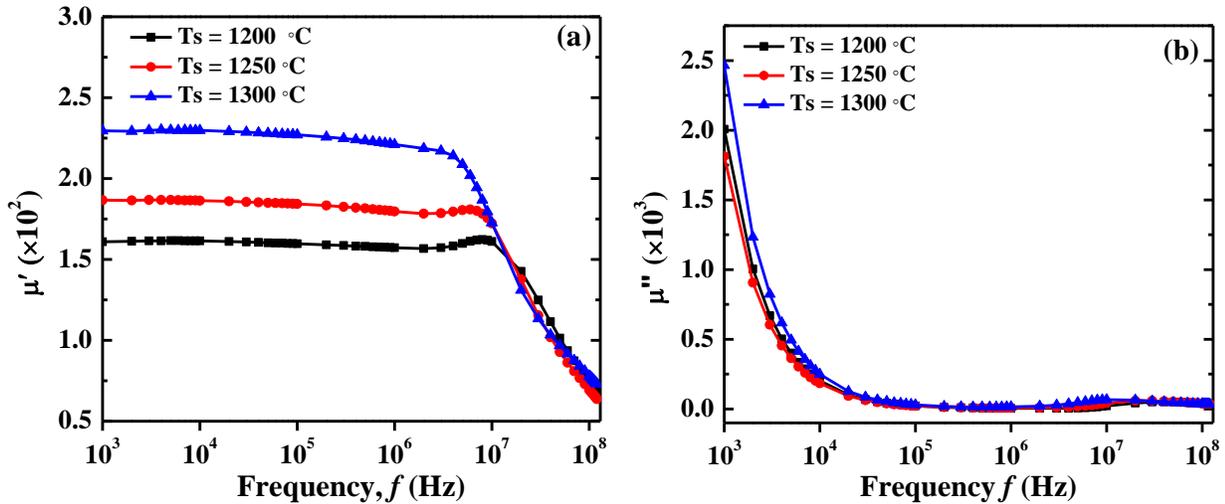

**Fig. 3.** The frequency dependence of permeability (a) real part (b) imaginary part of $Ni_{0.6}Zn_{0.4}Fe_2O_4$ samples sintered at 1200, 1250, 1300 °C for 4 hrs in air.

The loss component is the imaginary part of initial permeability ($\mu''$) (magnetization is $90°$ out of phase with the alternating magnetic field) and represents ultimate operating frequency of magnetic devices. Fig. 3 (b) depicts the loss component of the NZFO as a function of frequency. The value of $\mu''$ is much higher at lower frequencies. It decreases rapidly at lower frequencies but at high frequencies its value becomes so small that it becomes independent of frequency. It is also noticed that the loss component increases with increasing $T_s$. The grain size increases with the increase of $T_s$, as a result the number and size of magnetic domains rises which are contributing to loss due to delay in domain wall motion.

Frequency dependent relative quality factor or quality factor (Q=$\mu'$/tan$\delta$, $\mu'$ is the real part of initial permeability and tan$\delta$ is the loss factor of the samples) of the samples sintered at 1200, 1250, and 1300 °C is shown in Fig. 4. The high value of Q and $\mu'$ is expected for high frequency magnetic applications. The Q-factor increases with an increase of frequency showing a peak and decreases with further increase of frequency. It is seen that the Q-factor deteriorates beyond 10 MHz, i.e., the loss tangent is minimum up to 10 MHz and then it rises rapidly. The loss is due to the lag of domain wall motion with respect to the applied alternating magnetic field and is attributed to the various domain effects [24] such as non-uniform and non-repetitive domain wall motion, domain wall bowing, localized variation of flux density, nucleation and annihilation of domain walls. This happens at the frequency where the permeability begins to drop with frequency. This phenomenon is associated with the ferrimagnetic resonance within the domains and at the resonance maximum energy is transferred from the applied magnetic field to the lattice resulting in the rapid decrease in Q-factor. It is also seen that the Q value increases with decreasing $T_s$ and the maximum value found to be 6534 at $f = 2\times10^6$ MHz for the sample sintered at 1200 °C. The imperfection and defects in 1200 °C sintered sample are lower than that of samples sintered at higher $T_s$, which enhance the operating frequency range of hopping electrons between $Fe^{2+}$ and $Fe^{3+}$. Smaller grain size is competent for larger Q values, due to this reason, some grades of Ni–Zn ferrites are deliberately sintered at lowtemperature.

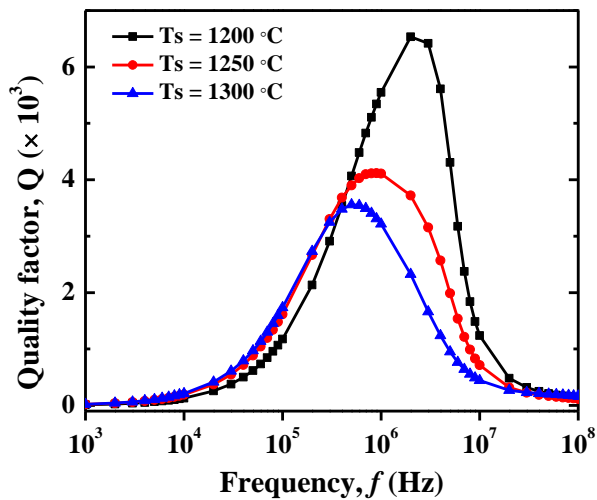

**Fig. 4.** The variation of Q factor with frequency of $Ni_{0.6}Zn_{0.4}Fe_2O_4$ samples sintered at 1200, 1250 and 1300 °C for 4 hrs in air.

Curie temperature ($T_c$) is the transition temperature above which the ferrite material loses its magnetic properties. Temperature dependence of initial permeability, $\mu_i$ of the toroid shaped of NZFO at constant frequency 1MHz of an AC signal is shown in Fig. 5. With the increase of temperature, the initial permeability increases rapidly and then drops off sharply near the transition temperature known as $T_c$ showing the Hopkinson effect [25]. The value of $K_1$ becomes almost negligible near the $T_c$ results the peak is obtained. At $T_c$, complete spin disorder takes place, i.e., a ferromagnetic material converts to a ferromagnetic material. Moreover, the sharpness of the permeability drop during phase transition indicates the compositional homogeneity and the verification of single phase cubic spinel of the studied samples, which has also been confirmed by X- ray diffraction (Fig.1).

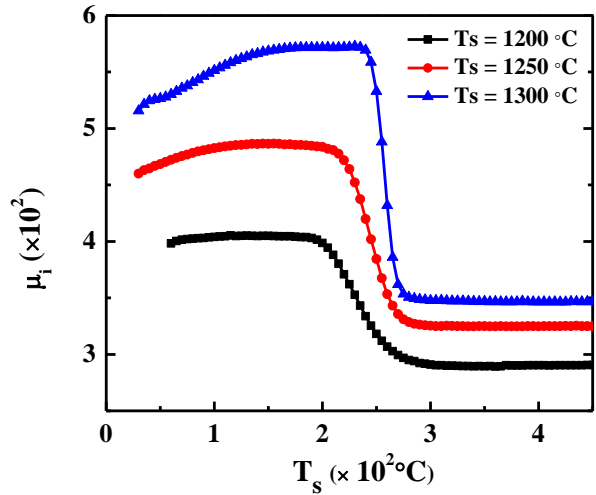

**Fig. 5.** The temperature dependence of the initial permeability ($\mu_i$) of $Ni_{0.6}Zn_{0.4}Fe_2O_4$ ferrite samples sintered 1200, 1250 and 1300 °C for 4 hrs in air.

## 4. Conclusions

The effect of sintering temperature on the structural and magnetic properties of $Ni_{0.6}Zn_{0.4}Fe_2O_4$ has been investigated. The parameters such as grain size, saturation magnetization and coercive field have been measured and well compared with the available data. The grain size of $Ni_{0.6}Zn_{0.4}Fe_2O_4$ synthesized from nano powders as raw materials are smaller than that of synthesized from bulk powders. Wide range of operating frequency or stability region of the $Ni_{0.6}Zn_{0.4}Fe_2O_4$ 1 kHz to 10 MHz has been determined which is an advantage of that material. A very low $H_c$ (<3 Oe) implies that this material is a promising candidate for transformer core and inductor applications. The magnetic properties of the $Ni_{0.6}Zn_{0.4}Fe_2O_4$ synthesized from nano powders are also better than that of the Ni-Zn ferrite synthesized from bulk materials. It is expected that our study would stimulate the scientist to synthesis the materials/compounds with nano powders as raw materials.

**Acknowledgements**

We are grateful to the authority of CUET for the financial support of this work.